\documentclass[%
 rsi,
 amsmath,amssymb,
 reprint,%
]{revtex4-1}

\usepackage{graphicx}
\usepackage{dcolumn}
\usepackage{bm}

\usepackage[utf8]{inputenc}
\usepackage[T1]{fontenc}
\usepackage{mathptmx}
\usepackage{amsmath}
\usepackage{etoolbox}
\usepackage{comment}
\usepackage[dvipsnames]{xcolor}
\usepackage{ amssymb }
\usepackage[caption=false]{subfig}
\usepackage[title]{appendix}
\usepackage{enumerate}
\usepackage{physics}
\usepackage{url}
\graphicspath{{media/}}
\makeatletter
\def\@email#1#2{%
 \endgroup
 \patchcmd{\titleblock@produce}
  {\frontmatter@RRAPformat}
  {\frontmatter@RRAPformat{\produce@RRAP{*#1\href{mailto:#2}{#2}}}\frontmatter@RRAPformat}
  {}{}
}%
\makeatother
\begin{document}
\preprint{AIP/123-QED}

\title{The QICK Box: A Modular RF Front-End System for Quantum Control and Readout \\[1ex] \small The QICK development team}



 \author{L. Arnaldi$^1$}
 \author{D. Martin$^1$}
 \author{S. Sussman$^1$}
 \author{S. Uemura$^1$}
 \author{K. Treptow$^1$}
 \author{N. Wilcer$^1$}
 \author{A. Col{\'o}n Cesan{\'i}$^2$}
 \author{O. Seidel$^1$}
 \author{K. Anyang$^{1,3}$}
 \author{G. Bratrud$^{1,2}$}
 \author{F. Abouzahr$^{1,2}$}
 \author{J. Christiansen-Salameh$^{1,4}$}
 \author{D. Baxter$^{1,2}$}
 \author{G. Cancelo$^1$*}
 \email{cancelo@fnal.gov}
 
\affiliation{$^1$Fermi National Accelerator Laboratory, Batavia, IL 60510, USA}
\affiliation{$^2$ Department of Physics \& Astronomy, Northwestern University, Evanston, IL 60208, USA}
\affiliation{$^3$Department of Physics, Illinois Institute of Technology, Chicago, IL 60616, USA}
\affiliation{$^4$Sibley School of Mechanical and Aerospace Engineering, Cornell University, Ithaca, NY 14853, USA}

\date{\today}

\begin{abstract}
The QICK (Quantum Instrumentation Control Kit), first introduced in 2022, is a
 standalone open source control system for quantum information science. Since
	then, QICK has been migrated to new hardware platforms, expanded its
	functionality for applications with different types of qubits
	(superconducting, atomic, ion-trapped, and spin defects), as well as quantum
	sensors, and quantum networks. QICK is a highly optimized control kit, with a
	library of functions that can be used for more than one application. In
	particular, the software and firmware are customized to minimize latency and
	maximize control performance and throughput.  In this white paper, we describe hardware details and performance of the QICK Box. The QICK Box features up to 16 outputs, up to 8 inputs,
and 8 20-bit DC biases. Modular output and input daughtercards give the
	choice of signal chains optimized for fast DC-coupled signals or RF signals up
	to 10\,GHz.  Power levels, step attenuation, and bandpass filtering are
	controlled by software. We demonstrate the system's performance through
	single-shot readout measurements of a six-transmon device at the Quantum
	Underground Instrumentation Experimental Testbed (QUIET) at Fermilab,
	achieving $(91.7 \pm 0.4)\%$ readout fidelity limited by T$_1$ decay during the readout. The
	single shot readout signal-to-noise ratio (SNR) is $4.4 \pm 0.1$ without a
	quantum amplifier, confirming that the QICK Box is suitable for high-fidelity
	quantum control and measurement applications. Advances in firmware and
	software will be described in a separate paper.
\end{abstract}

\maketitle

\section{Introduction}\label{Intro}
The Quantum Instrumentation Control Kit (QICK) is an open-source,
FPGA-based platform for pulse generation, readout, and real-time control
of quantum computing, networking, and sensing
experiments~\cite{Stefanazzi2022, QICKrepo}.
The QICK takes advantage of a commercial platform hosting a Gen3 RFSoC FPGA
(ZCU216) from AMD~\cite{rfsocwebsite}, which integrates 16 high-speed
DACs, 16 ADCs, programmable logic, and a 4-core ARM microprocessor in the same
package. This new generation of RF DACs operates at sampling rates that are high enough 
to directly synthesize microwave pulses without any kind of analog upconversion~\cite{Kalfus2021}, eliminating the need for meticulous calibration~\cite{Jolin2020}.

\begin{figure*}[!]
     \centering
     \includegraphics[width=\textwidth]{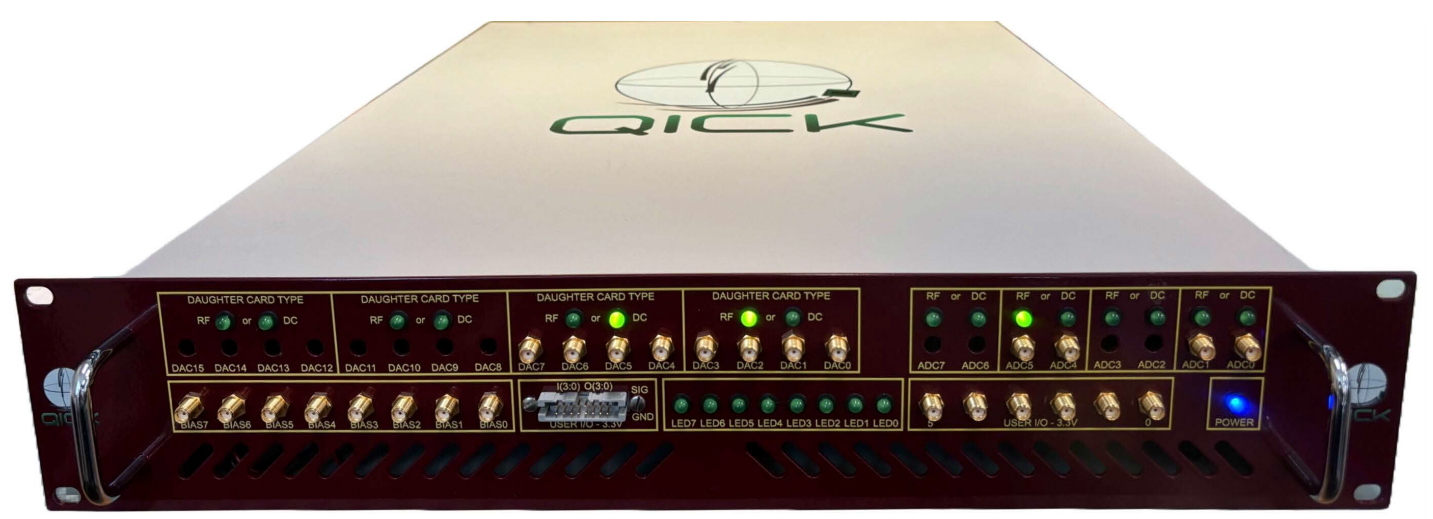}
		 \caption{\textbf{QICK Box 2U front-panel assembly.} The internal RF Main
		 board hosts the modular daughter cards, which are connected to the
		 front-panel SMA interfaces via high-performance coaxial cables. This 2U
		 rack-mount configuration provides a shielded environment for the RFSoC and
		 signal-conditioning electronics, ensuring robust connectivity for
		 large-scale quantum experiments.}
    \label{fig:QICKv2_outside}
\end{figure*}

The QICK firmware and software running on the ZCU216, hereafter referred to
as the QICK board, has been extensively validated across a range of quantum
information science experiments~\cite{Knirck2023, Xie2023, Riendeau2023, Ding2024, Stifter2024,
Feder2024, Bland2025, Hao2025, Zhang2025Yb, 
Shinozaki2025}. In
this white paper, we introduce a companion board for the QICK board with RF
and fast DC inputs and outputs to eliminate the need for extra discrete
electronics. The companion board includes eight 20-bit DACs for precision
biasing of devices and 16 digital inputs and outputs for markers and
triggers. While the use of the companion board is optional, it removes the
need for external room-temperature electronics, which is particularly
convenient when a large channel count is required. The QICK board together
with its companion board make up the QICK Box
(Fig.~\ref{fig:QICKv2_outside}). The companion board is the primary new
hardware contribution described in this paper.

\begin{figure*}[!htb]
     \centering
     \includegraphics[width=0.75\textwidth]{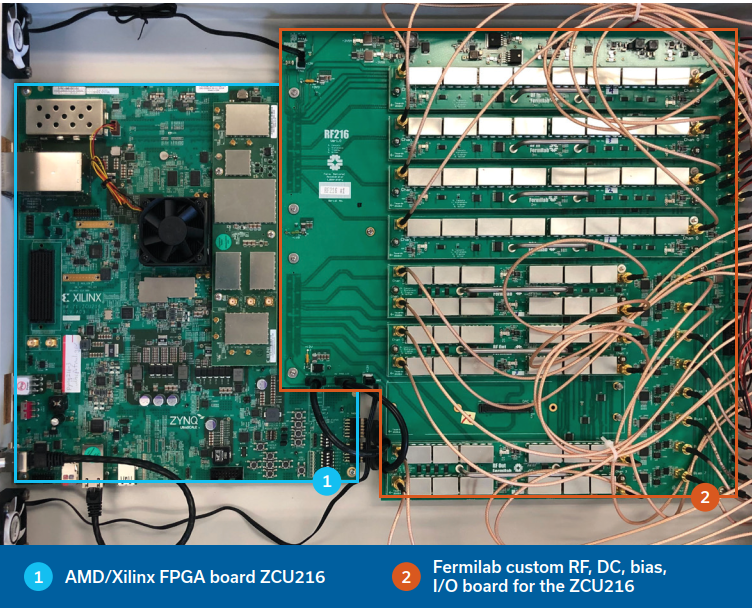}
		 \caption{\textbf{System integration of the QICK Box.} The architecture is
		 split between the AMD ZCU216 RFSoC (left), which performs digital
		 synthesis and high-speed data conversion, and the companion RF Main board
		 (right).  The RF Main board hosts the modular daughter cards responsible
		 for analog signal conditioning, including low-noise amplification,
		 programmable filtering, and power leveling across the
		 $0.1$--$9\,\text{GHz}$ range.}
    \label{fig:QICKv2_inside}
\end{figure*}

\section{The QICK Box}\label{sec:hardware}

\subsection{Analog Front-End Architecture}
The analog front-end of the QICK companion board serves as the critical
interface between the ZCU216's RF-sampling Data Converters (RF-DCs) and the
quantum experiment. To maintain high dynamic range and minimize phase noise, the
front-end is implemented using a modular motherboard-daughtercard architecture. 
 
\begin{itemize} 
	\item[1.] \textit{Signal Conditioning Chain}

	The signal path is designed to optimize the SNR (Signal-to-Noise Ratio) across a wide bandwidth (100\,MHz to 9\,GHz). Each channel utilizes a
	multi-stage approach:
\begin{itemize}
	\item Amplification: Dual-stage gain using PMA3-83LN+ amplifiers provides the
		necessary overhead for driving cryogenic attenuators.
	\item Precision Attenuation: PE43705 digital step attenuators allow for
		0.25\,dB resolution, enabling precise calibration of Rabi oscillations and
		gate pulses across the 0.1-9\,GHz spectrum.
	\item Filtering: The ADMV8818 digitally programmable bandpass filter acts as a
		tunable ``window,'' suppressing high-order harmonics and images from adjacent
		Nyquist zones.
\end{itemize}

\item[2.] \textit{Clock and Phase Stability, and Deterministic Low-Latency Communication}

Synchronous operation and communication across all FPGAs in a multi-board setup are achieved via the XCOM network~(\cite{XCOM2026}).

\begin{itemize}
    \item The XCOM network allows up to 16 QICK boxes to be connected in a mesh topology.
    \item The XCOM protocol phase-synchronizes the PLLs on all boards with a resolution better than 50\,ps.
    \item Any board can send peer-to-peer or broadcast messages with a deterministic latency of 95\,ns per 32-bit word. Supported payload sizes are 8, 16, and 32 bits.
\end{itemize}

\item[3.] \textit{DC Biasing (Auxiliary Front-End)}

Beyond RF signaling, the front-end integrates high-precision DC biasing. Incorporating 20-bit DACs, the system provides ultra-stable voltage offsets ($\pm$10\,V) for tuning qubit frequencies and controlling flux-tunable couplers. Maintaining a low noise floor (7.8\,nV/$\sqrt{\text{Hz}}$ with a $1/f$ knee at 4\,Hz) is essential to minimize dephasing in the quantum processors.

\item[4.] \textit{Digital I/O}

Sixteen general-purpose digital I/O ports provide low-latency pathways for hardware-level synchronization, external triggering, and the real-time control of peripheral components.

\end{itemize}

\subsection{ZCU216 companion board}\label{AC_DC}

The companion board is architected as a motherboard supporting four distinct
daughter card configurations (Fig.~\ref{fig:QICKv2_inside}). Acting as an
extension of the ZCU216, the motherboard interfaces directly via two RFMC
connectors, from which it also draws power. Consequently, the QICK Box requires
only a single external 12\,V DC (100\,W minimum) power supply connected to the
ZCU216. Beyond power management, the motherboard handles addressing,
programming, and channel enabling for the daughter cards.  Additionally, it
hosts eight high-precision DC biasing channels based on 20-bit DACs. These bias
outputs provide a $\pm$10\,V dynamic range with 1\,ppm resolution, 1\,ppm
integral nonlinearity, and a noise profile characterized by
7.8\,nV/$\sqrt{\text{Hz}}$ white Gaussian noise with a $1/f$ knee at 4\,Hz.

\begin{figure}
     \centering
     \includegraphics[width=0.49\textwidth]{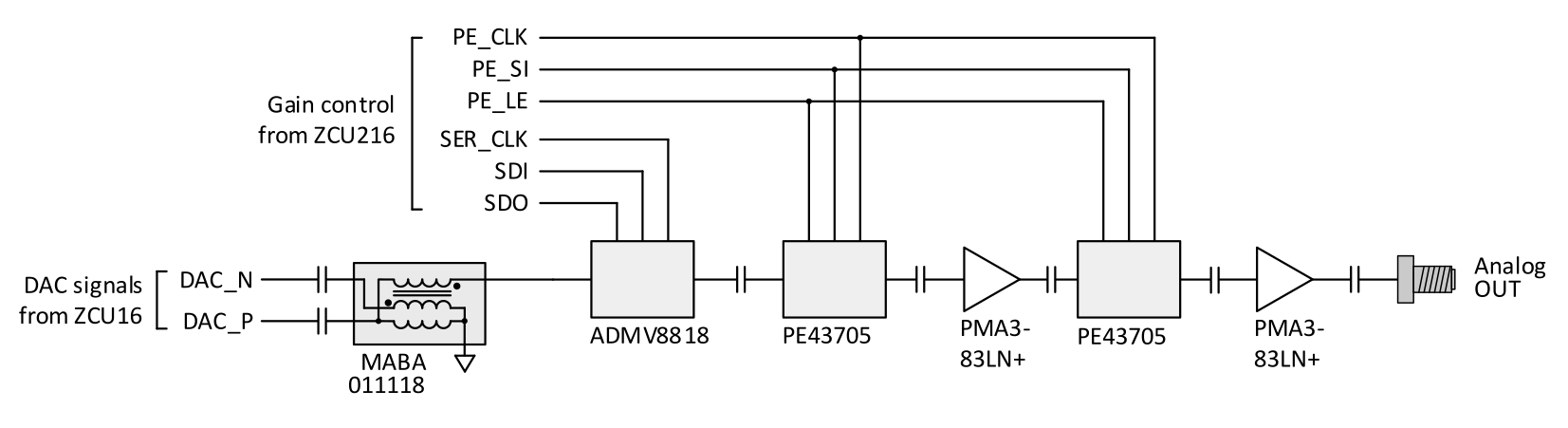}
		 \caption{\textbf{RF output daughter card block diagram.} The chain utilizes
		 a 10\,GHz balun for differential-to-single-ended conversion, followed by
		 an ADMV8818 digitally programmable bandpass filter to suppress Nyquist
		 images and out-of-band spurs across the 100\,MHz to 9\,GHz range. The final
		 stage comprises two PMA3-83LN+ gain stages and two PE43705 digital step
		 attenuators.}
    \label{fig:DC_RFout}
\end{figure}

\begin{figure}
     \centering
     \includegraphics[width=0.49\textwidth]{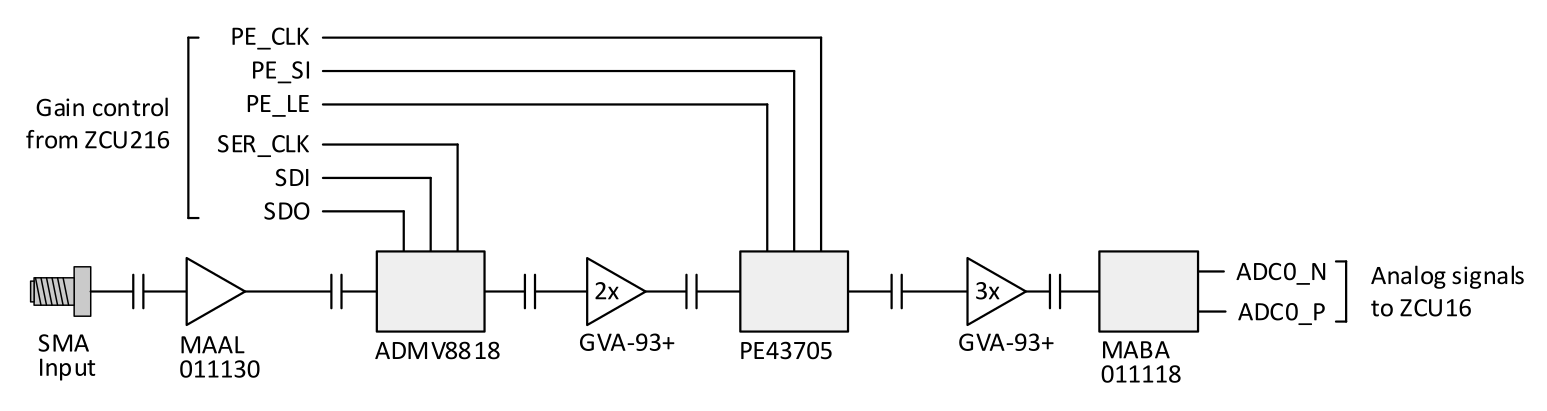}
		 \caption{\textbf{RF input daughter card block diagram.} The receiver chain
		 is optimized for high sensitivity, featuring a MAAL-011130 LNA as the
		 primary stage to achieve a $1.4\,\text{dB}$ noise figure. Signal
		 conditioning is managed via pSemi digital step attenuators and an Analog
		 Devices ADMV8818 tunable bandpass filter, providing up to $90\,\text{dB}$
		 of total gain and robust out-of-band rejection for quantum readout
		 applications.}
    \label{fig:DC_RFin}
\end{figure}

\begin{figure}
     \centering
     \includegraphics[width=0.49\textwidth]{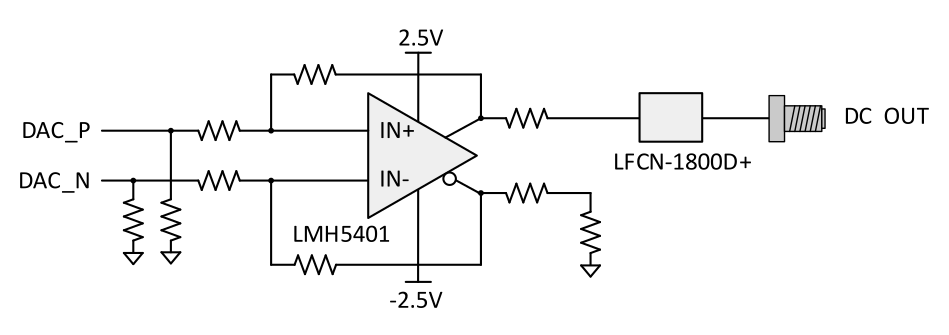}
		 \caption{\textbf{DC-coupled output daughter card block diagram.} The
		 architecture features a Texas Instruments LMH5401 in a single-ended
		 configuration with the negative output terminated to $50\,\Omega$. A
		 $1.8\,\text{GHz}$ low-pass filter (LPF) ensures suppression of
		 high-frequency images and clock feedthrough.}
    \label{fig:DC_DCout}
\end{figure}

\begin{figure}
     \centering
     \includegraphics[width=0.49\textwidth]{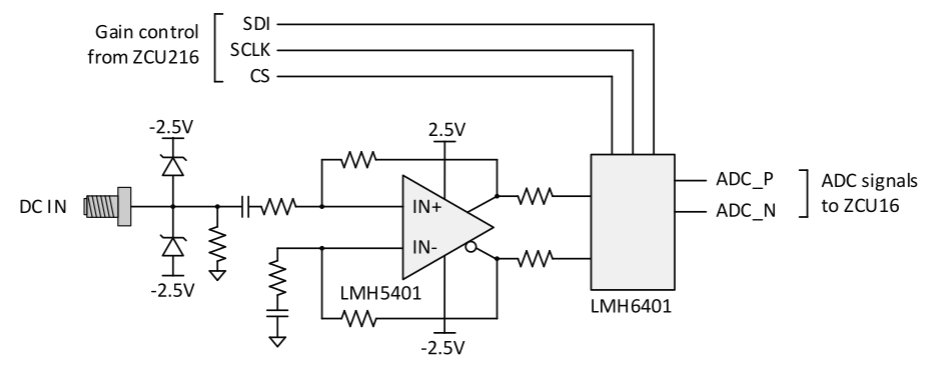}
		 \caption{\textbf{DC-coupled input daughter card block diagram.} The
		 architecture features a high-bandwidth two-stage amplification chain. The
		 first stage consists of an LMH5401 buffer ($4\,\text{GHz}$ BW), followed by
		 an LMH6401 digital variable gain amplifier ($4.5\,\text{GHz}$ BW). This
		 configuration provides up to $26\,\text{dB}$ of gain with $1\,\text{dB}$
		 resolution, supporting both high-speed quantum sensing and general-purpose
		 diagnostic functions.}
    \label{fig:DC_DCin}
\end{figure}

The daughter cards are available in six variants: RF outputs (100\,MHz--9\,GHz),
DC-coupled outputs (DC--1.5\,GHz), RF inputs (50\,MHz--9\,GHz), DC-coupled
inputs (DC--1.5\,GHz),  balun-only outputs, and balun-only inputs. Each output
daughter card supports four channels. The card flavor (RF or DC) is
automatically detected. The input daughter cards support two channels per card.
This lower input channel density is a deliberate design choice to align with
typical quantum system requirements, thereby reducing cost, power consumption,
and physical footprint. Detailed block diagrams for each daughter card variant
are provided in Figs.~\ref{fig:DC_RFout}--\ref{fig:DC_DCin}.

Each RF output channel (Fig.~\ref{fig:DC_RFout}) incorporates a 10\,GHz balun
followed by a signal conditioning chain comprising two 22.6\,dB gain amplifiers
(Mini-Circuits PMA3-83LN+), two 3--30\,dB programmable step attenuators (pSemi
PE43705), and a digitally programmable bandpass filter (Analog Devices
ADMV8818). The attenuators provide a fine tuning resolution of 0.25\,dB per step.
The inclusion of the programmable bandpass filter is critical for suppressing
out-of-band noise and eliminating spurious tones beyond the Nyquist zone of
interest, ensuring high signal fidelity for quantum control applications.

The RF input channels (Fig.~\ref{fig:DC_RFin}) provide a high-gain front-end
(90\,dB maximum) operating from 50\,MHz to 9\,GHz. The primary stage utilizes a
Macom MAAL-011130 Low-Noise Amplifier (LNA), achieving a noise figure of 1.4\,dB
(equivalent to a noise temperature of 110\,K). Similar to the RF output stage,
the input chain incorporates a pSemi PE43705 programmable step attenuator and an
ADMV8818 digitally tunable bandpass filter for precise gain control and
out-of-band interference rejection. The nominal input dynamic range is -95\,dBm
to -55\,dBm; however, signals below this range may be acquired with a
corresponding reduction in the Signal-to-Noise Ratio (SNR).

\begin{figure}
    \hfill
    \includegraphics[width=\columnwidth]{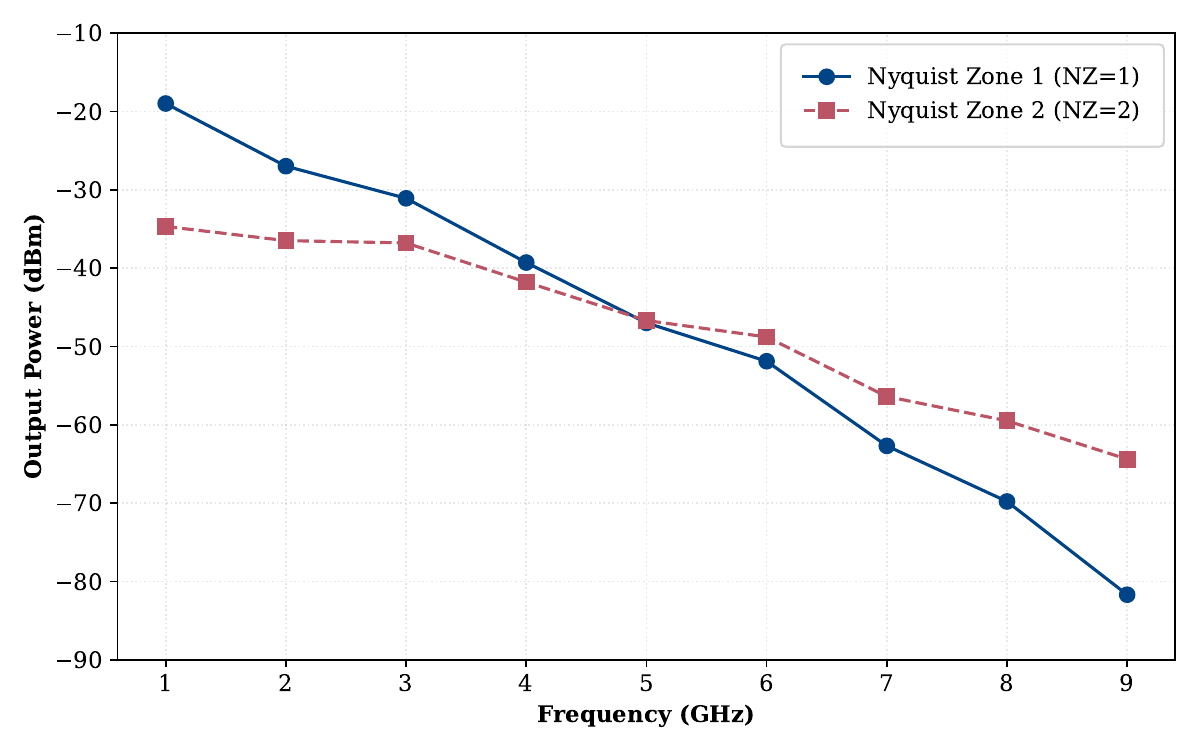}
		\caption{\textbf{DAC and RF output channel power vs. frequency.} The blue/solid
		trace shows the transfer function when the DAC Nyquist zone is set to 1, the
		red/dashed trace when the DAC Nyquist zone is set to 2. The daughtercard amplifier
		gains droop 2\,dB in that frequency range. The RF trace insertion loss is
		also larger for higher frequencies.}
    \label{fig:RFout}
\end{figure}

Each DC-coupled output channel (Fig.~\ref{fig:DC_DCout}) utilizes a Texas
Instruments LMH5401 fully differential amplifier configured for unity gain,
followed by a low-pass filter with a 1.8\,GHz cutoff frequency. Although the
LMH5401 is an ultra-wideband fully differential amplifier, the channel is
configured for single-ended, non-inverting output. The complementary negative
output is internally terminated with a 50\,$\Omega$ load to maintain impedance
matching and balance.

These DC-coupled channels are specifically designed to drive fast, unmodulated
signals, such as the high-speed flux pulses required for fluxonium qubits
\cite{Zhang2021}, or baseband voltage pulses for gate-controlled spin qubits in
quantum dots \cite{Hanson2007}.

The DC-coupled input channels (Fig.~\ref{fig:DC_DCin}) utilize a two-stage
amplification architecture designed for high-speed signal acquisition. The first
stage employs a Texas Instruments LMH5401 fully differential amplifier with
unity gain and a 4\,GHz bandwidth. This is followed by an LMH6401 digital
variable gain amplifier (DVGA) providing a 4.5\,GHz bandwidth and a maximum gain
of 26\,dB.  The LMH6401 features an integrated programmable attenuator with a
range of 0--30\,dB in 1\,dB increments, allowing for precise input scaling. The
input stage exhibits a noise figure of 8\,dB at 1\,GHz.

In quantum sensing, these DC-coupled inputs enable direct detector readout, with the QICK's Time Tagger IP providing picosecond-level timing resolution~\cite{Xie2023}. Concurrently, these channels serve as auxiliary diagnostic tools, functioning as an integrated oscilloscope or spectrum analyzer.

\begin{figure}[h!]
\begin{center}
			\includegraphics[width=\columnwidth]{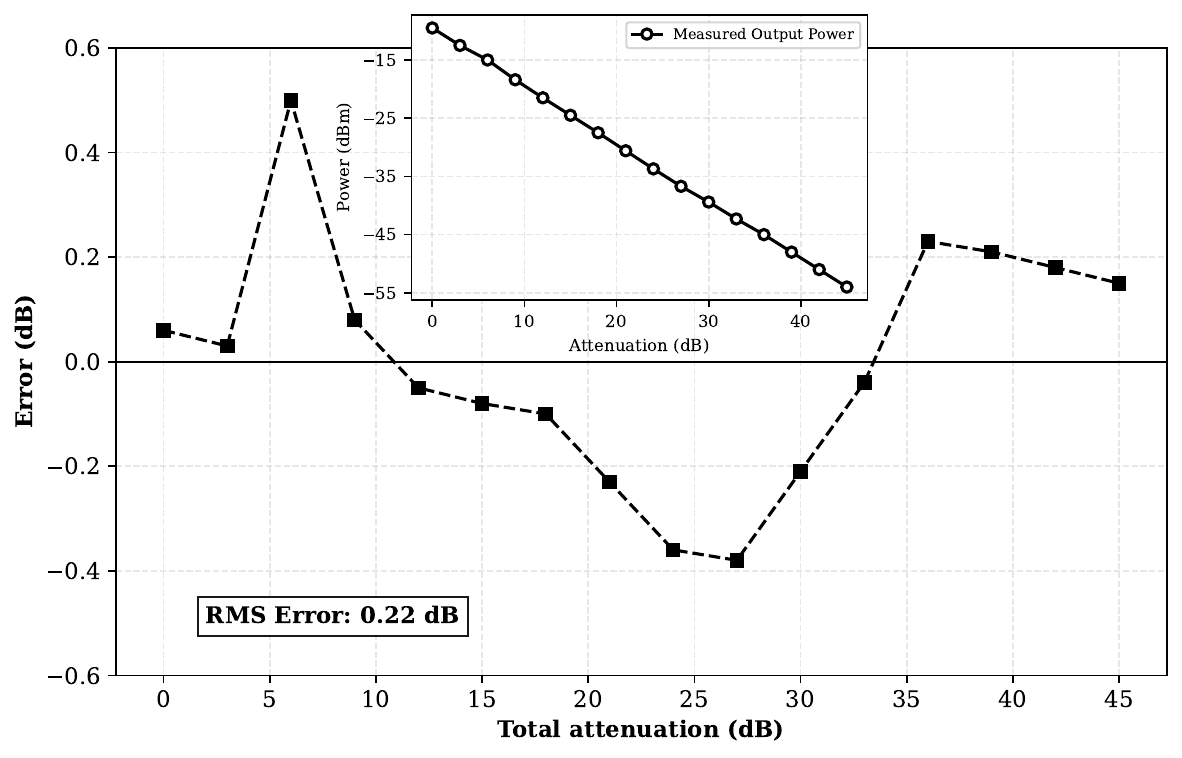}
			\caption{\textbf{RF output power linearity:} Measured output power ($P_{\text{out}}$) as a function of the programmed attenuation setting from $0$ to $45$\,dB (upper inset), and the deviation from the ideal linear response (residuals, lower plot), exhibiting a characterization error within $0.22$\,dB RMS. The high linearity across the full $1$--$9$\,GHz spectrum enables precise power equalization and reliable calibration for high-fidelity quantum control pulses.}
		\label{fig:RFout_vs_att}
\end{center}
\end{figure}

\subsection{AMD ZCU216 RFSoC evaluation board}\label{RFSOC_FPGA}

The QICK architecture leverages the high-level integration of the AMD XCZU49DR
RFSoC. This SoC features sixteen 14-bit, 10\,GSPS digital-to-analog converters
(DACs) and sixteen 12-bit, 2.5\,GSPS analog-to-digital converters (ADCs). These
converters interface directly with the FPGA fabric via standard AXI-Stream
protocols, eliminating the high-power consumption associated with external
JESD204B or LVDS interfaces. Both the DAC and ADC tiles incorporate
comprehensive digital signal processing (DSP) chains, including configurable I-Q
digital up/down-conversion (DUC/DDC), integrated numerically controlled
oscillators (NCOs), complex gain matrices, and interpolation/decimation filters.
Beyond the RF fabric, the device hosts a quad-core ARM Cortex-A53 processor for
executing quantum algorithms, 70\,Mb of on-chip memory, and diverse connectivity
interfaces.

By operating within the DAC's second Nyquist zone, the QICK system generates complex-envelope I-Q RF signals near 10\,GHz via direct digital upconversion. The QICK firmware supports multiple Signal Generator (SG) architectures capable of modulating envelopes up to 3\,GHz bandwidth onto a 0--10\,GHz carrier. This wide-bandwidth generation shifts harmonic images further from the fundamental frequency, significantly relaxing the requirements for analog reconstruction filters. Unlike traditional Arbitrary Waveform Generators (AWGs), QICK Signal Generators are highly memory-efficient, leveraging parametric synthesis to minimize the memory footprint (see~\cite{Stefanazzi2022, Ding2024} for implementation details).
\begin{figure}[h!]
    \centering
    \includegraphics[width=\columnwidth]{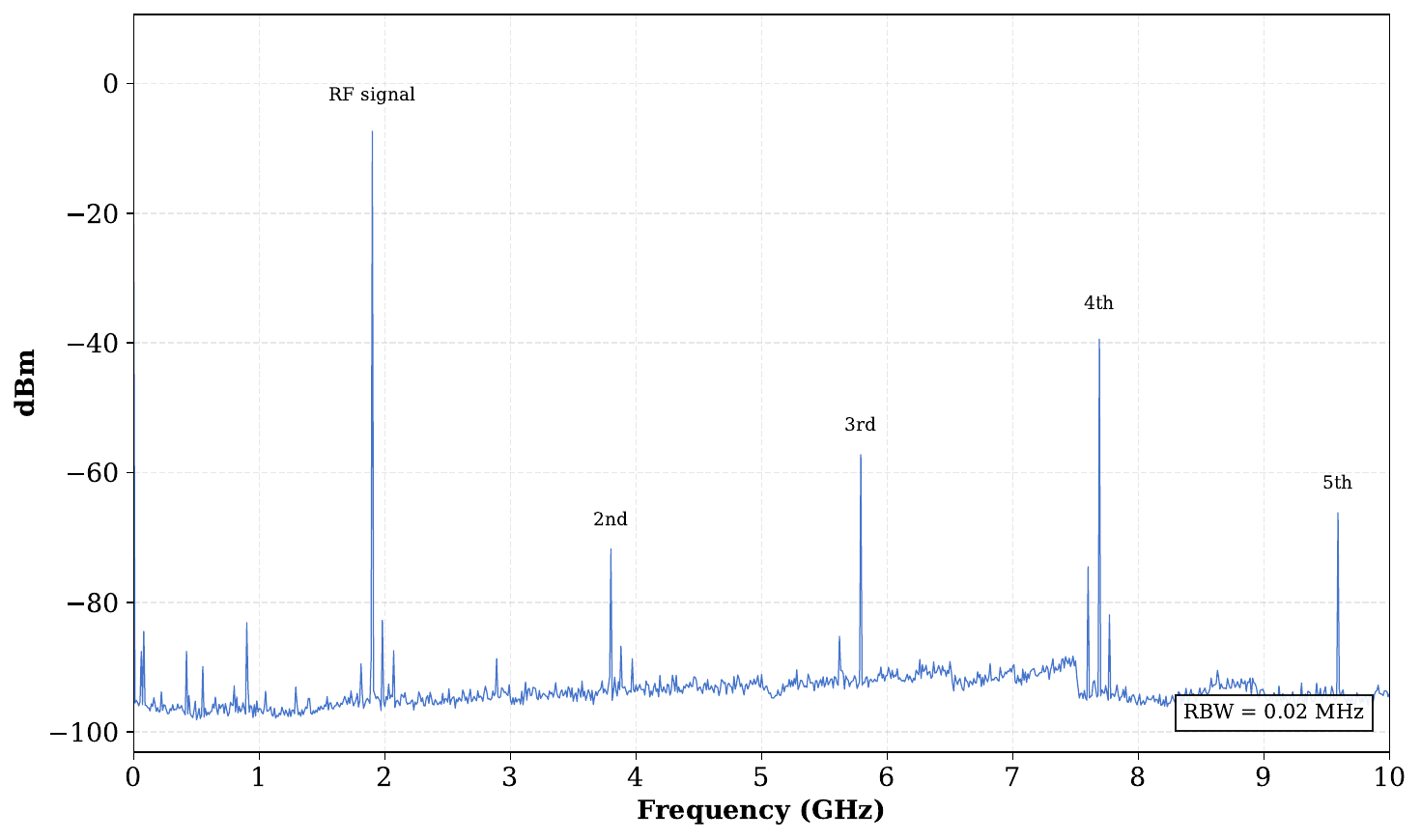}
    \caption{\textbf{RF output harmonic content.} Wideband spectrum for a 1.9\,GHz single-tone RF signal, showing the 2nd through 5th harmonic components. Excluding these higher-order harmonics, the spectrum exhibits a non-harmonic Spurious-Free Dynamic Range (SFDR) of approximately 75\,dBc (RBW = 0.02\,MHz).}
    \label{fig:DAC_PoutH}
\end{figure}
\subsection{QICK Box characterization}\label{QICK_char}

This section details the performance characterization of the QICK system
integrated with the RF companion board.
\begin{figure}[h!]
\begin{center}
\includegraphics[width= 1\columnwidth]{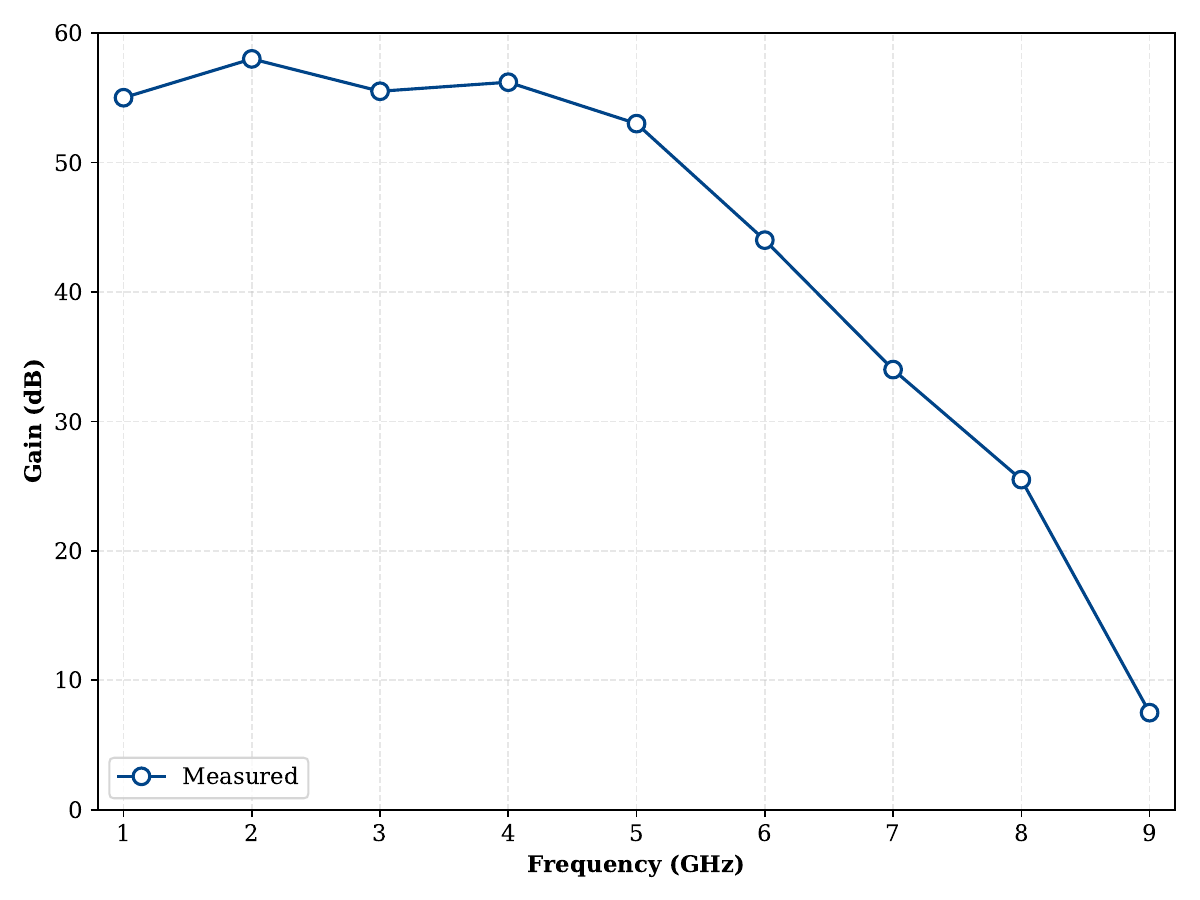}
\caption{\textbf{Measured RF input gain as a function of frequency.} The
	observed divergence at higher frequencies is primarily due to the analog
	bandwidth limitations of the ZCU216 ADCs and the insertion loss of the
	$11$-inch PCB traces. With the attenuator set to $30\,\text{dB}$, the system
	maintains high sensitivity across the $1$--$8\,\text{GHz}$ range, with a
	maximum attainable gain of $\approx 90\,\text{dB}$ when variable attenuation
	is removed.}
  \label{fig:RFin_gain}
\end{center}
\end{figure}

\subsubsection{RF Output Performance}\label{RF_out}
\begin{figure}[h!]
\begin{center}
\includegraphics[width= 1\columnwidth]{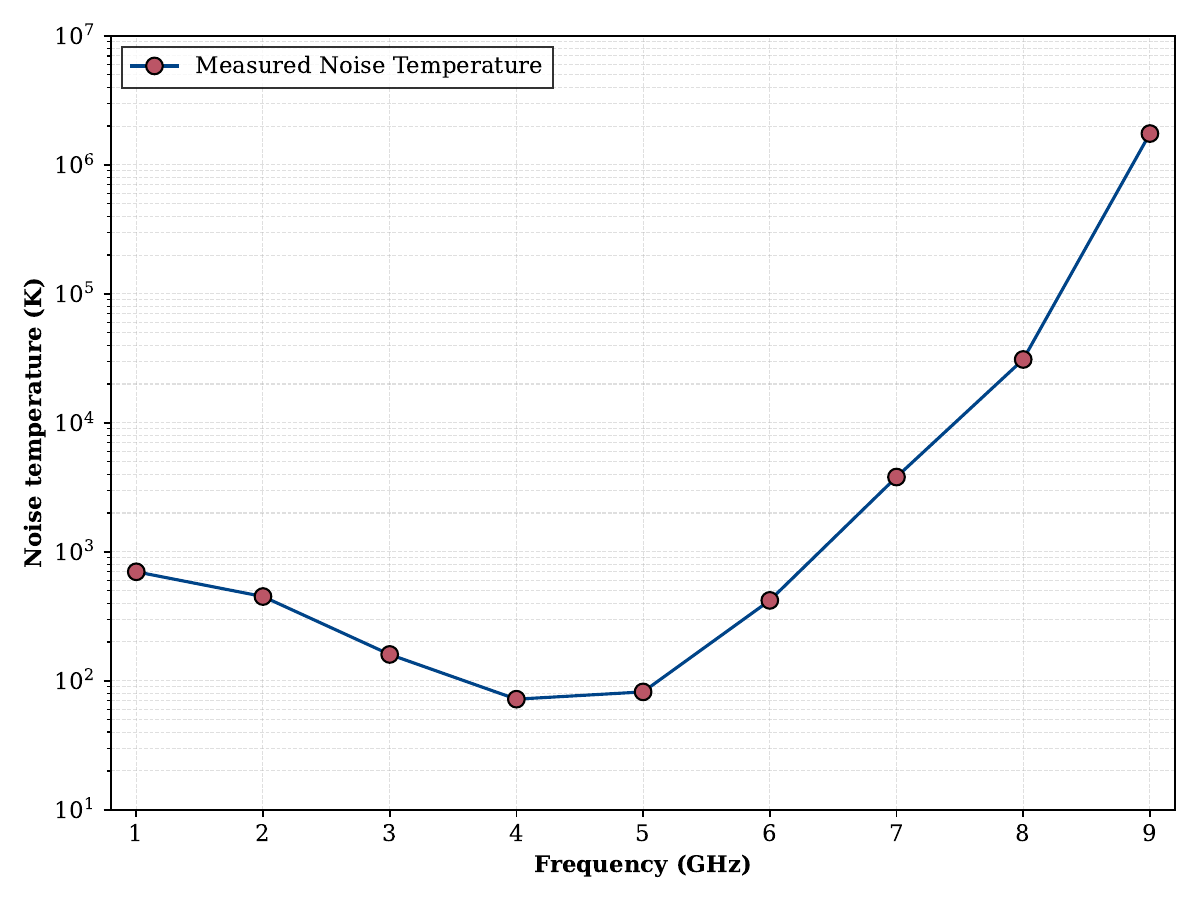}
\caption{\textbf{RF input noise temperature versus frequency.} The system
	maintains a noise temperature of $\approx\,100$\,K (corresponding to a
	$1.4$\,dB noise figure) within the $3$--$6$\,GHz band. The observed increase
	in noise temperature at higher frequencies is a result of the reduction in
	front-end gain; as the primary stage gain decreases, the noise contributions
	from subsequent components in the cascaded signal chain become more dominant.}
        \label{fig:RFin_noise}
\end{center}
\end{figure}
A randomly selected RF channel was utilized for baseline characterization;
inter-channel consistency results are detailed in subsequent sections.
Fig.~\ref{fig:RFout} illustrates the DAC output power ($P_{out}$) in dBm as a function of frequency from 1 to 9\,GHz with the attenuators set to $\text{Att}_1 = 20\,\text{dB}$ and $\text{Att}_2 = 30\,\text{dB}$ to prevent compression at lower frequencies.

\begin{itemize}
	\item \textbf{Nyquist Zone Optimization:} 
     The blue/solid traces represent DAC
		operation in the first Nyquist zone (NZ=1), while the red/dashed traces
		correspond to the second Nyquist zone (NZ=2). To maximize output power, the
		system should transition from NZ=1 to NZ=2 for frequencies at or above
		5\,GHz. The higher gain of the blue trace from 0.1--5\,GHz follows the gain of the DAC digital filters for Nyquist zone 1. For the interval 5--9\,GHz selecting the DAC in Nyquist zone 2 improves the gain with respect to Nyquist zone 1. The DAC Nyquist zone is a software parameter in QICK.
	\item \textbf{RF output power linearity} (Fig.~\ref{fig:RFout_vs_att}):    Measured output power ($P_{\text{out}}$) as a function of the programmed attenuation setting from $0$ to $45$\,dB (upper inset). Deviation from the ideal linear response (residuals), showing a characterization error within $0.22$\,dB RMS (lower plot). The high linearity across the full $1$--$9$\,GHz spectrum enables precise power  equalization and reliable calibration for high-fidelity quantum control pulses.
	\item \textbf{Power Output Frequency Response and Path Loss:} Some
		equalization can be achieved by choosing the proper DAC Nyquist zone. A
		$-2$\,dB power droop is observed at higher frequencies for each of the 2
		cascaded amplifiers. Frequency-dependent insertion loss in the RF
		traces also reduces gain at high frequencies (see Fig.~\ref{fig:RFout}). The
		step attenuators are used to equalize and flatten the output power within a
		$P_{out}$ range from -10\,dBm to -40\,dBm.
\end{itemize}

\paragraph{Spectral Purity and Noise} Excluding higher-order harmonic products, the RF output features a non-harmonic Spurious-Free Dynamic Range (SFDR) of approximately 75\,dBc, which is compatible with the specifications reported in the AMD RFSoC datasheet~\cite{ds926}. Fig.~\ref{fig:DAC_PoutH} illustrates the wideband spectrum for a 1.9\,GHz fundamental signal, detailing the 2nd through 5th harmonic components, which align with the expected device performance outlined in~\cite{ds926}.

\begin{itemize}
	\item \textbf{Filtering:} Both input and output daughter cards integrate a
		software-programmable chip combining low-pass (LP) and high-pass (HP)
		stages. When configured as a bandpass filter, the system achieves image
		rejection of 20\,dB at 1\,GHz from the carrier. On the input channels, this filter is used as a
		robust anti-aliasing and noise-rejection filter.

	\item \textbf{Noise Floor:} For a single-tone output of -20\,dBm at 8\,GHz,
		the measured noise floor 1\,GHz away from the carrier is below -150\,dBc/Hz.
\end{itemize}

\paragraph{Maximum Output Power and Saturation}

The maximum output power available before hitting compression is $\approx 10$\,dBm. At higher frequencies, the maximum available power decreases due to
the combined gain roll-off of the DAC, the RF amplifiers, and the PCB trace
attenuation. At 9\,GHz, the maximum output power is greater than $-10$\,dBm.

\subsubsection{RF Input Performance}\label{RF_in}
\begin{figure}[h!]
\begin{center}
\includegraphics[width= 1\columnwidth]{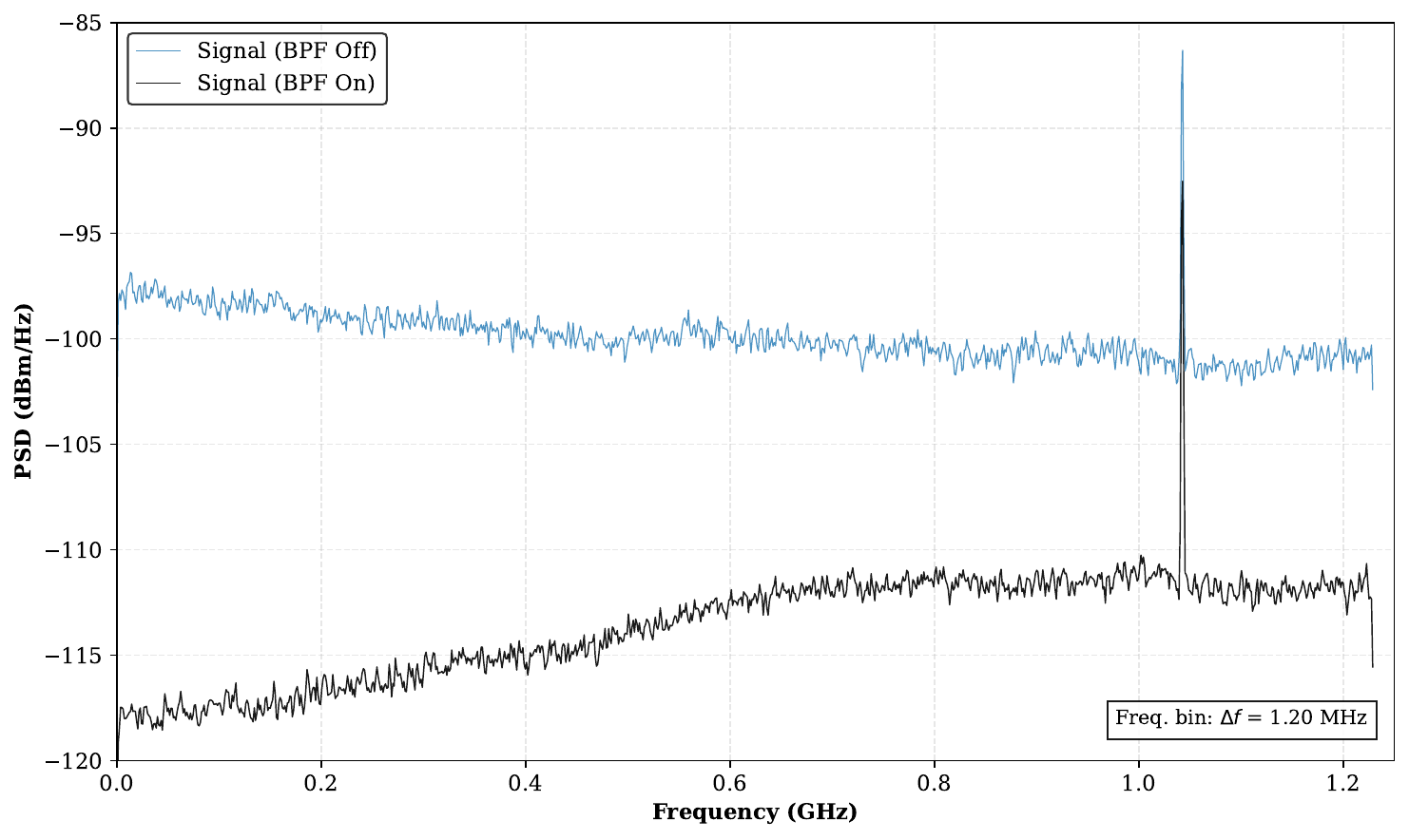}
\caption{\textbf{Effect of software-programmable BPF on input noise floor.}
		(a) PSD of a $-87$\,dBm signal and $50\,\Omega$ terminated noise floor with
		the BPF disabled. The wide analog bandwidth of the ADC allows noise from
		multiple Nyquist zones to alias into the signal band. (b) The same
		measurement with a $1$\,GHz BPF centered on the carrier. The filter reduces
		the integrated noise power by $\approx 10$\,dB at $f > 500$\,MHz from the carrier, demonstrating the necessity
		of band-limiting the input for high-fidelity quantum readout.}
\label{fig:RFin_filter}
\end{center}
\end{figure}
For the RF input characterization, a single channel was selected at random.
Fig.~\ref{fig:RFin_gain} displays the measured transfer function when the programmable step attenuator is set to its maximum value of 30\,dB. When the variable attenuation is removed (0\,dB), the system achieves a maximum gain of $\approx\,90\,\text{dB}$ within the 1--5\,GHz
range. The 30\,dB attenuator range provides sufficient headroom to equalize the
gain across the 1--8\,GHz spectrum. At 9\,GHz, however, the gain of both the ADC
and RF amplifiers drops significantly; at this frequency, even with zero
attenuation, the maximum system gain is limited to 37\,dB. Measurements were
performed using an input signal generated by an RF output channel and external
fixed attenuators to maintain a power level between -78 and -82\,dBm. In all
tests, the bandpass filter (BPF) was configured with a 1\,GHz bandwidth centered
on the carrier frequency.

The noise performance of the RF input daughter card was characterized using a
50\,$\Omega$ termination at room temperature. During measurements, the digital step attenuator
was fixed at 30\,dB, and the tunable bandpass filter (BPF) was swept from 1 to
9\,GHz with a constant 1\,GHz passband. The input-referred noise was calculated
using the system gain profiles established in the previous section.

As shown in Fig.~\ref{fig:RFin_noise}, the system maintains a noise temperature
of approximately 100\,K (corresponding to a noise figure of 1.4\,dB) between 3 and
6\,GHz. At frequencies above 6\,GHz, the noise temperature increases significantly.
This degradation is a direct result of the reduced gain in the primary RF
stages; as the first-stage gain drops, the total system noise figure becomes a
composite of the primary LNA noise and the noise contributions from subsequent
stages in the chain, as governed by Friis's formula for cascaded networks.

A key feature of the QICK RF hardware is the integration of
software-programmable bandpass filters (BPF). At the RF input, the filter is
typically centered on the readout carrier frequency with a half-bandwidth of $\approx $0.4\,GHz.  Given the ADC sampling rate of 2.4576\,GSPS, a wide-band signal chain would allow noise from multiple Nyquist zones to alias into the baseband, significantly degrading the Signal-to-Noise Ratio. The BPF effectively acts as a tracking anti-aliasing filter, suppressing out-of-band noise regardless of whether the primary signal resides in the first, second, or third Nyquist zone.

Fig.~\ref{fig:RFin_filter} displays the Power Spectral Density (PSD) of a
-87\,dBm input test signal in blue/solid when the filter is OFF and black/solid when the filter is ON. When set for a bandwidth of $\approx$0.4\,GHz the filter adds 10\,dB of insertion loss, which has been taken into account in the amplification chain. The benefit of the filter is evident in Fig.~\ref{fig:RFin_filter} lowering the noise floor by 20\,dB at 1\,GHz from the carrier.

\begin{figure}[h!]
\begin{center}
\includegraphics[width= 1\columnwidth]{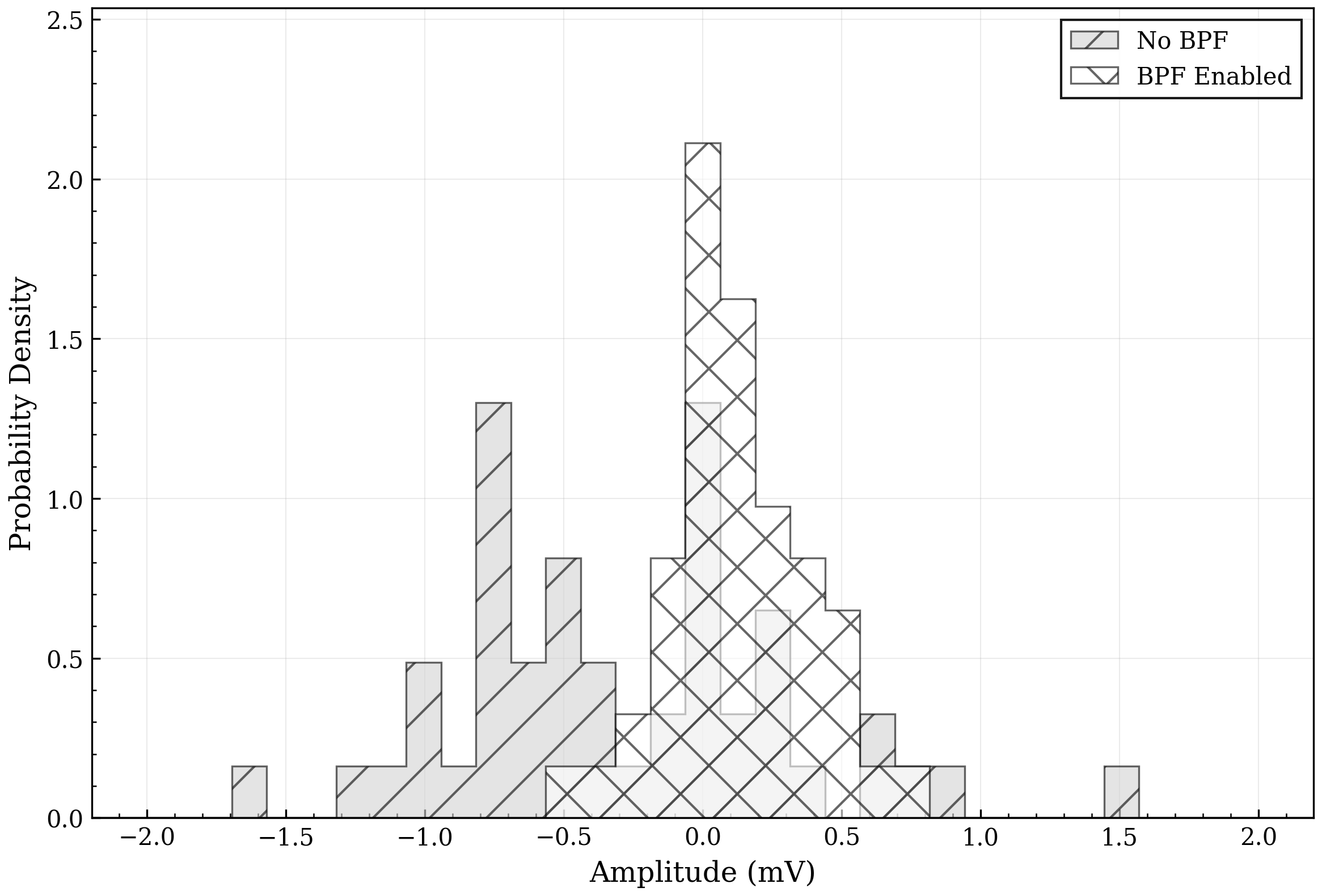}
\caption{\textbf{Impact of BPF on integrated noise statistics.} 
Histograms
of 50 noise datasets integrated over $1\,\mu$s ($2.64$\,K samples). The
distribution without BPF (diagonal hatch) shows significantly higher
variance due to aliased wide-band noise. The distribution with BPF
enabled (cross hatch) demonstrates a $7.6\,\text{dB}$ reduction in RMS noise,
from $0.596\,\text{mV}$ to $0.249\,\text{mV}$. This improvement in the integrated
signal-to-noise ratio (SNR) allows for higher-fidelity qubit readout
and faster state discrimination.
}
\label{fig:Rfin_BPF_noBPF}
\end{center}
\end{figure}

In a typical quantum experiment, qubit readout is performed by integrating 
the ADC-sampled signal over a window of several $\mu$s. To quantify the 
impact of the programmable bandpass filter (BPF) on readout fidelity, 
we compared integrated data sets with and without filtering. For this 
characterization, $132\times10^3$ samples were acquired and partitioned
into 50 segments of $2.64\times10^{3}$ points each. At a sampling
rate of $f_s = 2.4576\,\text{GSPS}$, each segment corresponds to an 
integration time of $\approx\,1\,\mu$s. Fig.~\ref{fig:Rfin_BPF_noBPF} displays 
the resulting histograms for these 50 datasets. The histogram without BPF
(diagonal hatch) represents the integrated noise floor, where the variance 
is broadened by aliased wide-band noise. The histogram with BPF enabled 
(cross hatch) shows the results with the $1\,\text{GHz}$ BPF activated. 
The implementation of the BPF as an anti-aliasing filter results in a 
$7.6\,\text{dB}$ reduction in RMS noise after 1\,$\mu$s of integration, 
reducing the integrated noise from $0.596\,\text{mV}$ to $0.249\,\text{mV}$. 
This reduction in noise variance directly translates to increased readout fidelity
and shorter required integration times for high-fidelity qubit readout.

\section{QICK Box measurements}

\subsection{Single-shot readout results}\label{sec:ssf_results}

\begin{figure*}[t!]
\begin{center}
\includegraphics[width= \textwidth]{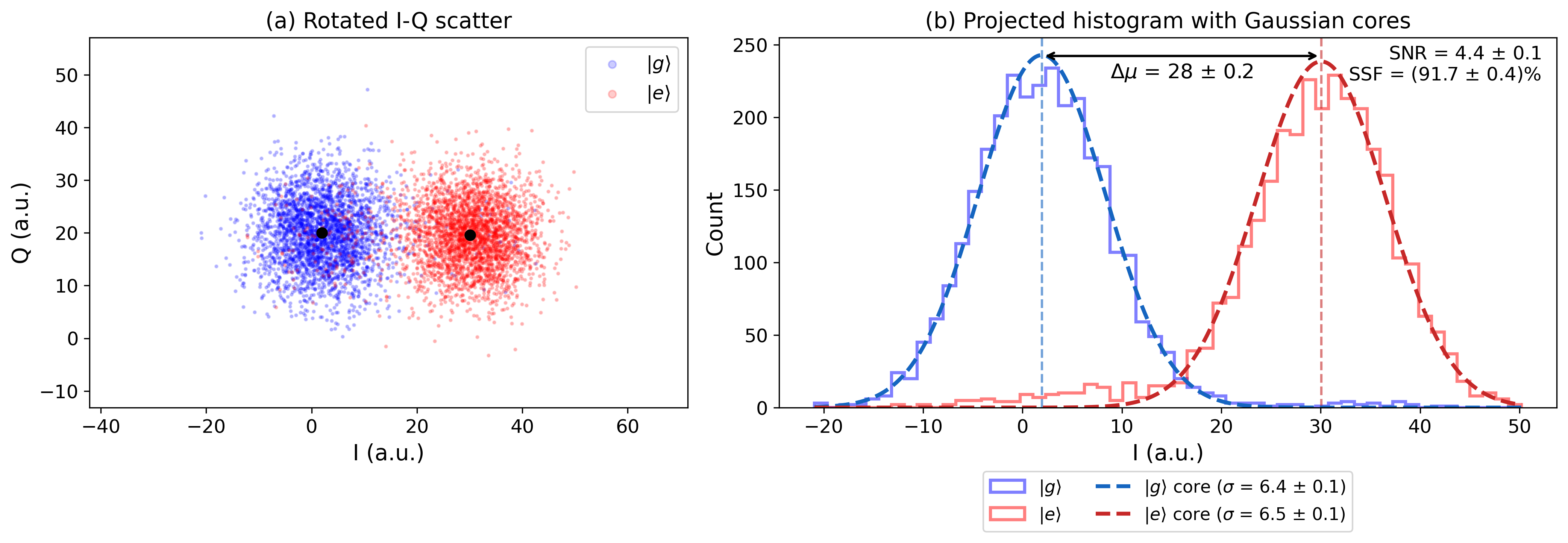}
\caption{\textbf{Single-shot qubit readout and state discrimination
using the QICK Box at QUIET.}
(a) Rotated I-Q scatter plot showing the ground (blue) and excited
(red) state clusters for 3000 single-shot measurements per state.
The data are rotated to align the axis of maximum state separation
with the I-quadrature. Black markers indicate the median of each
distribution.
(b) Projected histogram along the I-quadrature after rotation.
Dashed curves show Gaussian core fits obtained from the clean
(outer) side of each distribution, isolating the electronics noise
contribution from T$_1$ decay tails. The state separation is
$\Delta\mu = 28 \pm 0.2$\,a.u., with clean-side widths
$\sigma_g = 6.4 \pm 0.1$\,a.u.\ and
$\sigma_e = 6.5 \pm 0.1$\,a.u., yielding a readout electronics
SNR of $4.4 \pm 0.1$ and a readout fidelity of
$(91.7 \pm 0.4)\%$. The readout chain has a single HEMT
amplifier with no quantum amplifier.}
\label{fig:ssf_quiet}
\end{center}
\end{figure*}

We characterized the readout performance of the QICK Box through
single-shot dispersive measurements of a six-qubit tantalum-on-sapphire transmon device at the Quantum Underground Instrumentation Experimental
Testbed (QUIET) at Fermilab. The signal amplification chain before the QICK Box consisted of a single
High-Electron-Mobility Transistor (HEMT) amplifier at the cryogenic stage, with
no quantum amplifier (e.g.\ TWPA or JPA) in the signal path.
Despite this minimal amplification configuration, the system achieved a readout
electronics signal-to-noise ratio of
\begin{equation}
	\mathrm{SNR} = 4.4 \pm 0.1
\end{equation}
and a single-shot readout fidelity of
\begin{equation}
	F_{\mathrm{SSF}} = (91.7 \pm 0.4)\%\,,
\end{equation}
determined from 3000 single-shot measurements per qubit state.
The I-Q scatter plot and projected histogram with Gaussian core overlays are
shown in Fig.~\ref{fig:ssf_quiet}. 

The observed readout fidelity is primarily limited by T$_1$ decay during the
readout window: the device has a median relaxation time of
$T_1 = 58.3 \pm 4.1\,\mu$s, so a non-negligible fraction of the prepared
$\ket{e}$ population decays to $\ket{g}$ before state assignment, reducing the
contrast between the two clusters; a quantitative estimate is given in
Sec.~\ref{sec:fidelity_limits}. The qubit effective temperature of
$\approx\,44\,\text{mK}$ corresponds to a small residual thermal population of
$\ket{e}$ that accounts for a smaller share of the misclassification rate.
An ongoing study at QUIET is systematically adding infrared
shielding to the device to understand and model how each stage of filtering
affects coherence times and the qubit effective temperature~\cite{ColonCesani2026_QUIET}. Improvements in
$T_1$ are expected to increase the readout fidelity without
requiring changes to the QICK Box electronics.

\subsection{Experimental setup}

These measurements were conducted at the QUIET testbed, a 10\,mK dilution refrigerator located 100\,m underground
to mitigate cosmic-ray interference. QUIET provides the ultra-low-background environment required for
the precise characterization of superconducting qubits and high-coherence
cryogenic detectors.

Measurements were carried out on a six-qubit tantalum-on-sapphire transmon device with Manhattan-style Josephson junctions fabricated by our team at the Pritzker Nanofabrication Facility at the University of Chicago and packaged at Fermilab. Each fixed-frequency qubit is capacitively coupled to a dedicated readout resonator, and all resonators share a common feed-through line used for both dispersive readout and qubit drive. The qubit transition frequencies span 3.8--5.0\,GHz, detuned by approximately 2\,GHz from their corresponding readout resonators at 6.2--6.5\,GHz (linewidths $\approx$ 0.4--0.6\,MHz). All experiments addressed qubits individually; no entangling operations were performed.

\subsection{Single-shot readout analysis and SNR estimation}\label{sec:ssf_analysis}

Single-shot measurements serve as a primary diagnostic tool in superconducting
circuits, providing the necessary data to quantify state readout fidelity,
estimate thermal populations (effective temperature), and evaluate readout
contrast. Unlike averaged measurements, which obscure stochastic processes,
single-shot readout characterizes the qubit state in individual measurement
events.

The readout fidelity is fundamentally governed by the separation between the
ground ($\ket{g}$) and first excited ($\ket{e}$) state measurement outcomes,
visualized as distinct clusters in the complex I-Q plane
(Fig.~\ref{fig:ssf_quiet}). By applying a coordinate rotation to align the axis
of maximum state separation with the I-quadrature, the results can be projected
into a 1D histogram. The overlap between these two distributions determines the
single-shot readout fidelity and is limited by several mechanisms. These
include T$_1$ decay, drive-induced transitions, finite thermal population, and
electronic noise. Relaxation during the measurement window can lead to state
misclassification, placing a fundamental bound on the maximum achievable
fidelity, even in the ideal limit of noiseless amplifiers and electronics.
Imperfect readout or control pulses can also induce leakage to higher excited
states or introduce measurement back-action, effectively mixing populations
during acquisition. In addition, a nonzero thermal population means the qubit
does not begin purely in the ground state, reducing the contrast between the
prepared ground and excited-state clusters. Overall, single-shot fidelity
reflects both the performance of the measurement chain and the intrinsic
dynamics of the qubit itself. Higher SNR provided by the QICK input chain
results in greater separation and, consequently, enables active feedback
protocols such as real-time qubit reset.
\\
\paragraph{SNR estimation via clean-side Gaussian core fitting.}
Because T$_1$ decay during the readout window contaminates the inner flanks of
the $\ket{g}$ and $\ket{e}$ distributions, a direct standard-deviation estimate
would overstate the electronics noise. We therefore employ a clean-side fitting strategy: only the outer half of each distribution, i.e.\ the side unaffected by relaxation or thermal tails, is used to extract the
Gaussian core width. Specifically, the samples satisfying $I \leq
\widetilde{\mu}_g$ are used for the ground state and $I \geq
\widetilde{\mu}_e$ for the excited state, where $\widetilde{\mu}_{g,e}$ denote
the respective medians. This procedure isolates the contribution of the readout
electronics chain (amplifiers, ADC quantization, and thermal noise in the warm
signal path) from qubit thermalization and decoherence effects.

The readout electronics SNR is then defined as
\begin{equation}\label{eq:snr_def}
	\mathrm{SNR} = \frac{\Delta\mu}{\bar{\sigma}},
	\qquad
	\bar{\sigma} = \sqrt{\frac{\sigma_g^2 + \sigma_e^2}{2}}\,,
\end{equation}
where $\Delta\mu = |\widetilde{\mu}_e - \widetilde{\mu}_g|$ is the separation
between the median I-quadrature values and $\sigma_{g,e}$ are the clean-side
RMS widths.
\\
\paragraph{Statistical uncertainties.}
All quoted uncertainties are propagated from finite-sample statistics:
\begin{equation}\label{eq:err_mu}
	\delta\mu_{g,e} = \sigma_{g,e}\,\sqrt{\frac{\pi}{2\,N_{g,e}}}\,,
\end{equation}
\begin{equation}\label{eq:err_sigma}
	\delta\sigma_{g,e} = \frac{\sigma_{g,e}}{\sqrt{2\,N_{g,e}^{\text{clean}}}}\,,
\end{equation}
where $N_{g,e} = 3000$ is the number of shots per state and
$N_{g,e}^{\text{clean}} \approx 1500$ is the number of clean-side samples. The
uncertainty on the SNR follows from standard ratio error propagation:
\begin{equation}\label{eq:err_snr}
	\frac{\delta(\mathrm{SNR})}{\mathrm{SNR}}
	= \sqrt{
		\left(\frac{\delta(\Delta\mu)}{\Delta\mu}\right)^{2}
		+\left(\frac{\delta\bar{\sigma}}{\bar{\sigma}}\right)^{2}
	}\,.
\end{equation}
The single-shot readout fidelity is obtained from the maximum cumulative
histogram contrast, with a binomial uncertainty
$\delta F = \sqrt{F(1-F)/N_{\text{total}}}$.
\\
\paragraph{Results.}
The clean-side widths are $\sigma_g = 6.4 \pm
0.1$\,a.u.\ and $\sigma_e = 6.5 \pm 0.1$\,a.u., yielding an average noise
width of $\bar{\sigma} = 6.5 \pm 0.1$\,a.u. The state separation is
$\Delta\mu = 28 \pm 0.2$\,a.u.

The asymmetry between $\sigma_g$ and $\sigma_e$ (approximately 2\%) is
consistent with the expected contribution of residual T$_1$ decay leaking into
the nominally clean side of the excited-state distribution, and does not
indicate an electronics-related degradation of the input channel.

\subsection{Fidelity-limiting mechanisms}\label{sec:fidelity_limits}

The $\mathrm{SNR} = 4.4 \pm 0.1$ achieved using only a HEMT amplifier, with no
quantum-limited amplifier in the readout chain, demonstrates that the warm
electronics of the QICK Box are suitable for high-fidelity qubit readout. The
readout fidelity of $(91.7 \pm 0.4)\%$ is primarily limited by T$_1$ decay
during the readout window. Because the assignment is made on the IQ trace
averaged over the full readout pulse, a decay in the first half of the window
pulls the integrated point across the discrimination threshold and is
misclassified as $\ket{g}$, whereas a decay in the second half leaves the
integrated point on the $\ket{e}$ side of the threshold and is still correctly
assigned. With a readout pulse duration of $\tau_{\mathrm{ro}} = 5.6\,\mu$s and
a median qubit relaxation time of $T_1 = 58.3 \pm 4.1\,\mu$s, the resulting
T$_1$-limited assignment error is
\begin{equation}\label{eq:t1_limit}
	\varepsilon_{T_1} = 1 - \exp\!\left(-\frac{\tau_{\mathrm{ro}}}{2\,T_1}\right)
	= (4.7 \pm 0.3)\%\,,
\end{equation}
accounting for approximately $57\%$ of the measured total infidelity
$1 - F_{\mathrm{SSF}} = 8.3\%$. The remaining $\approx 3.6\%$ is attributable
to the residual thermal population associated with the qubit effective
temperature of $\approx 44\,\text{mK}$, imperfect state-preparation pulses,
and Gaussian overlap between the $\ket{g}$ and $\ket{e}$ distributions.
These measurements were
performed on a system in which infrared shielding and filtering are being
incrementally added across successive cooldowns to reduce the effective
temperature and improve coherence~\cite{ColonCesani2026_QUIET}. We did not observe a change in $T_1$ or effective qubit temperature as a result of switching from the QICK board setup to the QICK Box setup.

\section{Conclusion}
In this white paper, we have presented the QICK Box, the latest evolution of the
Quantum Instrumentation Control Kit. By transitioning to a custom-integrated box
with a modular approach, the QICK Box provides a versatile platform featuring up to
16 outputs, up to 8 inputs, and 8 20-bit DC biases, specifically designed to meet the rigorous demands of superconducting, atomic, and spin-based quantum systems.

Experimental characterization confirms that the modular signal chains offer
high-performance operation up to 10\,GHz. The RF output daughtercards provide
precise control with a step attenuation resolution of 0.25\,dB and a measured
linearity within 0.22\,dB RMS, enabling the effective equalization of power
across the full working bands. On the acquisition side, we demonstrated that
utilizing the second Nyquist zone maintains a high gain of >45\,dB at 9\,GHz, with
an optimal noise temperature of approximately 100\,K in the mid-band.
Furthermore, the software-controlled bandpass filtering was shown to be
essential for maximizing the Spurious-Free Dynamic Range (SFDR) and protecting
the integrity of weak quantum signals.

The system's real-world performance was validated through single-shot readout
measurements of a six-transmon device, achieving an SNR of $4.4 \pm 0.1$ and
$(91.7 \pm 0.4)\%$ readout fidelity. The
readout fidelity is limited by T$_1$ decay during the readout window ($T_1 = 58.3 \pm 4.1\,\mu$s). These results confirm that the QICK Box
provides the noise performance required
for high-fidelity quantum control and measurement, furthering the mission of
providing open-source, high-performance tools for the quantum information
science community.

\section*{Acknowledgements}
The authors acknowledge D. Bowring, E. Figueroa-Feliciano, R. Khatiwada, R. Linehan, D. Temples, D. Schuster, and M.  Hatridge, for important conversations and feedback as QICK Box users. The authors thank Ziqian Li for designing the six-transmon device used in this work. 
This material is based upon work supported by the National Science Foundation Graduate Research Fellowship Program under Grant No. DGE-2234667. Any opinions, findings, and conclusions or recommendations expressed in this material are those of the authors and do not necessarily reflect the views of the National Science Foundation. 
O.S. was supported by a Graduate Research Instrumentation Award (GIRA) through the Coordination Panel for Advanced Detectors (CPAD) as funded by the  DOE Office of High Energy Physics. 
J.C.-S. was supported by the U.S. Department of Energy, Office of Science, Office of Workforce Development for Teachers and Scientists, Office of Science Graduate Student Research (SCGSR) program. The SCGSR program is administered by the Oak Ridge Institute for Science and Education for the DOE under contract number DE-SC0014664. 
This manuscript has been authored by Fermi Forward Discovery Group, LLC,
acting under Contract No.\ 89243024CSC000002 with the U.S.\ Department of
Energy, Office of Science, Office of High Energy Physics. This work was
supported by the U.S.\ Department of Energy, Office of Science, National
Quantum Information Science Research Centers, Quantum Science Center, and
the U.S.\ Department of Energy, Office of Science, High-Energy Physics
Program Office. 

\bibliography{bibliography1.bib}

\end{document}